\documentstyle[12pt]{article}
\def\bg#1{\hbox{\bf#1}}
\font\twelve=cmbx10 at 15pt
\font\ten=cmbx10 at 12pt

\begin{document}

\begin{titlepage}

\begin{center}

{\ten Centre de Physique Th\'eorique - CNRS - Luminy, Case 907}

{\ten F-13288 Marseille Cedex 9 - France }

{\ten Unit\'e Propre de Recherche 7061}

\vspace{1 cm}

{\twelve NEMATOGENIC LATTICE MODELS IN ONE OR TWO DIMENSIONS
AND WITH LONG-RANGE INTERACTIONS}

\vspace{0.3 cm}

{\bf N. ANGELESCU}\footnote{Dept. of
Theoretical Physics, Institute for Atomic Physics, PO Box MG-6
Bucharest (ROMANIA)}{\bf, S. ROMANO}\footnote{Physics Dept.,
University of Pavia, via A. Bassi 6,
I-27100 Pavia (ITALY)} {\bf and V.A. ZAGREBNOV}\footnote{and D\'epartement de
Physique, Universit\'e d'Aix-Marseille II}

\vspace{1.5 cm}

{\bf Abstract}

\end{center}

Extending previous rigorous results, we prove existence of an
ordering transition at finite temperature for a class of nematogenic
lattice models, where spins are associated with a one- or
two-dimensional lattice, and interact via long-range potentials.

\vspace{3 cm}

\noindent February 1994

\noindent CPT-94/P.3012
\bigskip

\noindent anonymous ftp or gopher: cpt.univ-mrs.fr

\end{titlepage}

\section*{~~~}
%
Over the last twenty years, a number of rigorous results have been
obtained \cite{r01,r02,r03,r04,r05,r06,r07}, concerning existence
(or absence) of phase transitions at finite temperature in classical
lattice spin models with isotropic, $O(n)$ symmetric, interactions of
ferromagnetic character ($n \ge 2$, i.e. continuous spins); for one-
or two-dimensional models,
an ordering transition taking place at finite temperature can only
be produced by
appropriate long-range interactions \cite{r01}. Reflection
positivity and chessboard estimates have played a major
role in the relevant demonstrations \cite{r01}.

In the present paper, we extend
the named results
{}from ferromagnetic interactions to even functions of the scalar
product, thus making contact with lattice models for nematic liquid
crystals.

{}~~

We consider here a general classical lattice-gas (LG) model:
$\Lambda_d = Z^d$ denotes the lattice, ${\bg x}_k$ denotes the
dimensionless coordinates of the $k-$th lattice site, and $\nu_k
\in \{ 0,1 \}$ its occupation number. Occupied lattice sites host
$n-$component unit vectors
${\bg u}_k$, where $n=2,3$; the Hamiltonian thus reads
\begin{eqnarray}
H = \sum_{j < k} \nu_j W_{jk} \nu_k - \mu \sum_{k} \nu_k \\
W_{jk} = -F(|{\bg x}_j - {\bg x}_k|)
G({\bg u}_j \cdot {\bg u}_k);
\end{eqnarray}
here $\mu$ denotes the chemical potential, and the interaction
possesses $O(n)$ (or higher) symmetry. As a special case, the
corresponding
lattice spin (LS) model is defined by $\nu_k=1,~\forall k$, and
here the chemical potential becomes irrelevant.

We assume the function $F(R)$ to be non-negative at all distances,
and to possess reflection positivity. Furthermore, let $\Phi ({\bg
q})$ denote its lattice Fourier transform, and let the functional
\begin{eqnarray}
\Psi(F) =
\int \frac{d {\bg q}}{\Phi({\bf 0})-\Phi({\bg q})} \approx
\sum_{BZ} \frac{1}{\Phi({\bf 0})-\Phi({\bg q})} \end{eqnarray}
converge, (here $BZ$ denotes the first Brillouin Zone);
as for the function $G(\tau)$, we assume \begin{eqnarray}
G(\tau)=\left\{
\begin{array}{ll}
\tau & ,~n=2,3 \\
P_2(\tau)&,~n=3
\end{array}
\right.,
\end{eqnarray}
where $P_L$ are Legendre polynomials.

{}~~

An interaction $F(R)$ restricted at nearest neighbours ($n-n$) is
reflection-positive; its associated functional $\Psi(F_{n-n})$
converges when $d=3$ and diverges for $d=1,2$, and the same
dependence on dimensionality holds for any finite-ranged
interaction \cite{r08}.

Long-range interactions of the form
\begin{eqnarray}
F(R) = \epsilon R^{-p},~\epsilon > 0,~p > d , \end{eqnarray}
have been extensively studied in one and two dimensions; they
possess reflection positivity, and their functional $\Psi(F)$
converges for $d < p < 2d$, and diverges for $p \ge 2d$ \cite{r09}.

{}~~

For LS models and $G(\tau)=\tau$, the existence of an ordering
transition at finite temperature has been proven rigorously, under
the conditions stated above \cite{r01}; the existence of a
Berezhinskii-Kosterlitz-Thouless transition was also proven
rigorously for $d=2,~n=2$ and $n-n$ interactions \cite{r10}.
For $d=1,2$, inverse-power models defined by $p \ge 2d$ produce
orientational disorder at all finite temperatures
\cite{r04,r05,r06,r07};
available correlation
inequalities \cite{r11,r12,r13,r14,r15}
entail the existence of a
Berezhinskii-Kosterlitz-Thouless-like
transition for inverse-power interactions defined by $d=2,~n=2,~p
\ge 4$ \cite{r16};
this is also likely to happen
for $d=1,~n=2,3~,p=2$, and for $d=2,~n=3,~p=4$
\cite{r11,r17,r18,r19,r20,r21}.

The existence of an ordering transition at sufficiently large
chemical potential and finite temperature was proven rigorously
for the LG model defined by
$d=3,~n=3,~G(\tau)=P_2(\tau)$ and $n-n$ interactions \cite{raz};
its LS counterpart possesses an ordering transition at finite
temperature.

{}~~

Under the stated conditions on $F(R)$,
i.e. positivity, reflection positivity and convergence of
$\Psi(F)$, LG models defined by $G(\tau)=P_2(\tau)$ possess an
ordering transition for sufficiently large values of $\mu$ and
finite temperatures; their LS counterparts possess ordering
transitions at finite temperatures.

{}~~

The proof developed in Ref.\ \cite{raz}
for a three-dimensional model with $n-n$ interactions carries
through verbatim.

{}~~

This entails existence of an ordering transition at finite
temperature for one- and two-dimensional LS models with long-range
interactions of the inverse-power form and $d < p < 2d$.

These results were somehow implicit in various papers, but, to the
best of our knowledge, nowhere explicitly stated.

\end{document}